# Engineering of Ferroic Orders in Thin Films by Anionic Substitution

A. C. Garcia-Castro[1,*], Yanjun Ma[2,*], Zachary Romestan[2], Eric Bousquet[3], Cheng Cen[2] and Aldo H. Romero[2]

[1]School of Physics, Universidad Industrial de Santander, Calle 09 Carrera 27, 680002, Bucaramanga, Colombia.

[2]Department of Physics and Astronomy, West Virginia University, Morgantown, West Virginia 26506, USA.

[3]Physique Théorique des Matériaux, QMAT, CESAM, Université de Liège, B-4000 Sart-Tilman, Belgium.

**Abstract:**

Multiferroics are a unique class of materials where magnetic and ferroelectric orders coexist. The research on multiferroics contributes significantly to the fundamental understanding of the strong correlations between different material degrees of freedom and provides an energy-efficient route toward the electrical control of magnetism. While multiple $ABO_3$ oxide perovskites have been identified as being multiferroic, their magnetoelectric coupling strength is often weak, necessitating the material search in different compounds. Here, we report the observation of room temperature multiferroic orders in multi-anion $SrNbO_{3-x}N_x$ thin films. In these samples, the multi-anion state enables the room-temperature ferromagnetic ordering of the Nb $d$-electrons. Simultaneously, we find ferroelectric responses that originate from the structural symmetry breaking associated with both the off-center displacements of Nb and the geometric displacements of Sr, depending on the relative O-N arrangements within the Nb-centered octahedra. Our findings not only diversify the available multiferroic material pool but also demonstrate a new multiferroism design strategy via multi-anion engineering.

**Introduction:**

Ferromagnetism and ferroelectricity are essential for modern electronic technologies, ranging from advanced computing to sensing. The quest for materials where these two properties are intimately connected (i.e. multiferroic and magnetoelectric materials) has become an important effort to advance technological efficiency and fundamental understanding[1]. It is widely known that magnetic ordering often arises from the *d*-electrons of transition metal elements, but *d*-orbital occupancy has the tendency to suppress ferroelectric distortions[2]. This dilemma reflects the competing factors in generating multiferroism in bulk materials. The most widely exploited strategy to solve this dilemma is to combine separate functionalities from the A- and B-site cations. For example, *d*-electrons from transition metals that occupy the B-site can drive ferromagnetism while the A-site cations can provide the necessary driving force for structural distortions[2-5] due to either an anisotropic distribution of unbound valence electrons or size-dependent space-filling effects. This approach has successfully led to the discovery of most of the single-phase multiferroic materials identified to date, such as $BiFeO_3$[6], $YMnO_3$[7], $LuFe_2O_4$[8], and $ABF_3$ fluoride compounds [9,10,11].

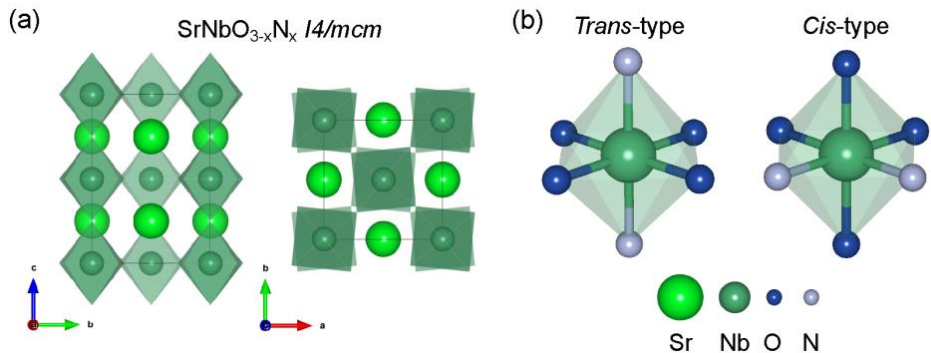

*Figure 1: Lattice structure of $SrNbO_{3-x}N_x$. (a) I4/mcm structure of the $SrNbO_{3-x}N_x$, here, out-of-phase octahedral rotation around the z-axis can be appreciated. (b) Depending on the relative coordinates of the nearby N atoms, oxynitride structures can be categorized into trans-type and cis-type.*

Instead of focusing on the cations, ferroelectricity and magnetism can also be stabilized by introducing a multi-anion state in perovskite oxides. To illustrate, replacing oxide ligands in the coordination octahedra by nitrogen creates an $ABO_{3-x}N_x$ oxynitride. The nitrogen substitutions in this example result in two possible anion configurations: the two nitrogen ions can occupy either adjacent (*cis*-type) or opposite (*trans*-type) sites in the $BO_4N_2$ octahedron as shown Figure 1. Both arrangements break local symmetry which can drive polar distortions. For instance, the *trans*-type ordering may drive the off-center displacement of Ta ions that cause ferroelectricity in $BaTaO_2N$[12]. Furthermore, density functional theory (DFT) calculations have shown that ferroelectricity can exist in both *trans-* and *cis-*type $LaTiO_2N$ thin films stabilized by compressive and tensile strain, respectively, having different saturation polarization and coercive fields[13]. In addition, the multiple anions introduce a mixed oxidation state which can also promote magnetic ordering. Colossal magnetoresistance has been revealed in $EuWO_{1+x}N_{2-x}$ structures[14], demonstrating this principle as well as the possibility for magnetoelectric coupling.

Although the *cis-*type coordination maximizes the overlap between the nitrogen $2p$ and transition metal $d$ orbitals and is energetically preferred over the *trans-*type configuration[15-17], most bulk oxynitrides do not show long-range anion order. The coexistence of both *cis-* and *trans-*type structures provides a new dimensionality that we can use to engineer multiferroism. In this work, we have synthesized $SrNbO_{3-x}N_x$ thin films epitaxially, characterizing them both experimentally and theoretically. The magnetic and ferroelectric properties of the thin films studied are found to be sensitively dependent on the nitrogen content. Additionally, the epitaxial strain also plays an important role in stabilizing and enhancing the ferroelectric ordering. As such, our results demonstrate multi-anion enabled room-temperature multiferroism for the first time.

**Results and Discussion:**

***Epitaxial growth of $SrNbO_{3-x}N_x$ thin films:***

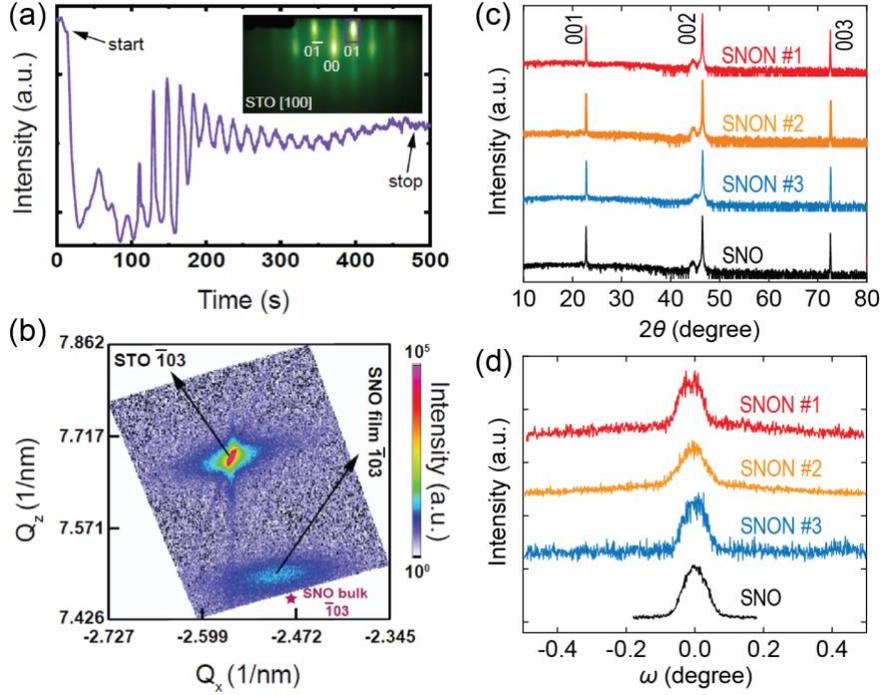

*Figure 2: Structural characterizations of SrNbO$_{3-x}$N$_x$ films. (a) RHEED oscillations observed during the growth of SrNbO$_3$ layer on top of STO (001) substrate, indicating a layer-by-layer growth mode. (b) RSM diagram of the SrNbO$_3$ film. (b) Symmetric XRD curves obtained on SrNbO$_3$ and SrNbO$_{3-x}$N$_x$ films. (d) Rocking curves obtained on SrNbO$_3$ and SrNbO$_{3-x}$N$_x$ films.*

First, 30 nm SrNbO$_3$ (SNO) films were grown on SrTiO$_3$ (STO) (001) single crystal substrates by pulsed laser deposition (PLD). The synthesis process was monitored *in-situ* by Reflection High Energy Electron Diffraction (RHEED) with the grazing incidence of the high energy electron beam along the STO [100] direction. The temporal changes in the 01 peak intensity during the deposition are shown in Figure 2a. Clear RHEED oscillations are observed, indicating a layer-by-layer growth mode. At the early stage of growth, the oscillation signal exhibits signatures of frequency doubling. Similar phenomena have also been observed in the growth of Ge[18], LaFeO$_3$[19] and SrTiO$_3$[19,20] films, which can be attributed to the inelastic scattering by periodic surface roughing. Similar to what is described by the step density model, the diffraction intensity first decreases when island formation roughens the surface and then increases when the islands coalesce[20-23]. Therefore, the presence of the frequency doubling in the first few oscillation cycles indicates that the interface between SNO and STO substrate may not be atomically sharp.

The crystalline quality of the SNO films was further studied by X-ray diffraction (XRD) (Fig. 2b, c). As shown by the reciprocal space mapping (RSM) (Fig. 2b), the SNO film grown is not completely commensurate to the substrate and experiences a small compressive strain. The in-plane and out-of-plane lattice constants of the SNO film are found to be $a \approx 4.00$ Å and $c \approx 4.02$ Å. In comparison, bulk SNO has a reported pseudocubic lattice constant of 4.023 Å[24].

Nitrogen was introduced into this oxide structure by annealing the grown SNO films in $NH_3$ gas (See Methods). Using this method, the nitrogen concentration in the resultant $SrNbO_{3-x}N_x$ (SNON) films can be tuned by controlling the sample temperature, $NH_3$ flux, and annealing time. Figure 2c and 2d show the $\theta$-$2\theta$ XRD scans and the rocking curves acquired from three samples annealed with different $NH_3$ fluxes (SNON #1, SNON #2, and SNON #3). Among them, SNON #1 corresponds to the most flux, and SNON #3 corresponds to the least. Each of the annealed SNON samples exhibits comparable crystallinity to the unannealed SNO film.

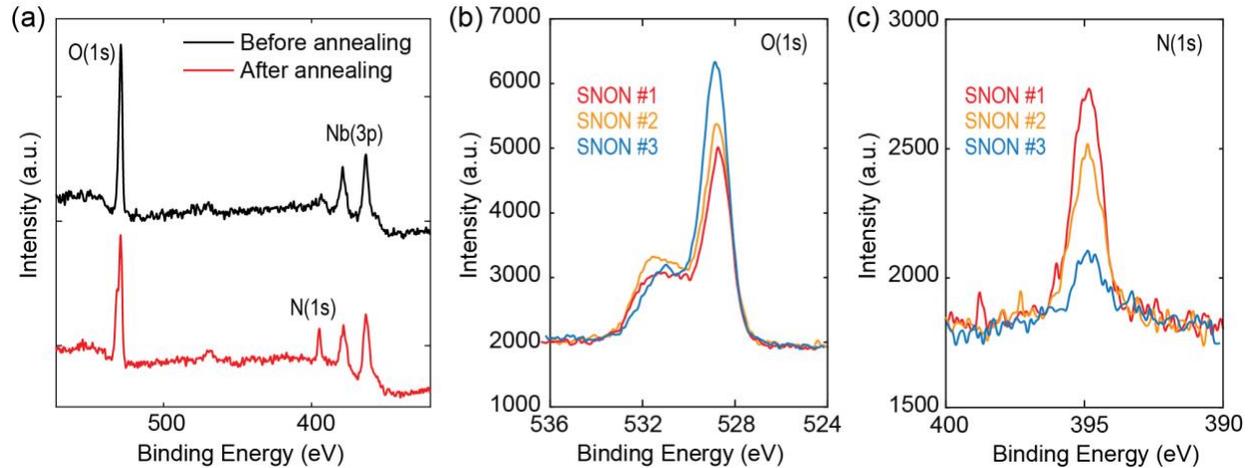

*Figure 3 Composition characterizations of SrNbO$_{3-x}$N$_x$ films. (a) XPS spectra measured on the same sample (SNON #3) before and after the NH$_3$ annealing. (b, c) Oxygen 1s and nitrogen 1s XPS peaks measured on the three SrNbO$_{3-x}$N$_x$ films.*

The nitrogen concentration in the annealed films was quantified by X-ray photoemission spectroscopy (XPS) measurements. As shown in Figure 3a, annealing the samples in $NH_3$ gas significantly affects the XPS characteristics associated with the anions. After annealing, a nitrogen 1s peak emerges near 395 eV, and the oxygen 1s peak near 530 eV splits into two. While the first feature is directly linked to the incorporation of nitrogen, the second feature is linked to it implicitly. Due to the different valences between $O^{2-}$ and $N^{3-}$, the introduction of N inevitably changes the effective valence of oxygen in the lattice, leading to the splitting of the O (1s) peak. XPS curves of the three annealed samples near the O (1s) and N (1s) peaks are compared in Figure 3b and 3c. As the NH$_3$ flux used during the annealing process increases (from SNON#3 to SNO#1), the N concentration increases, and the O concentration reduces. This observation is consistent with the substitution of O by N at the anion sites. After calibrations considering the XPS signals coming from all four elements, the compositions of the three $SrNbO_{3-x}N_x$ samples are found to be: $x$=0.3 (SNON #3), $x$=0.4 (SNON #2), and $x$=0.6 (SNON #1).

*Magnetic properties:*

In SrNbO$_3$, Nb$^{4+}$ has the electronic configuration $4d^1$, resulting in the intrinsic conductivity discussed previously in the literature[24]. After introducing nitrogen into the structure, the Nb cation could oxidize from Nb$^{4+}$ to Nb$^{5+}$ due to the difference in the oxidation state between $O^{2-}$ and the $N^{3-}$ substitution. The introduction of an anion with a different oxidation state has a profound effect on the magnetic properties, as revealed by the vibration sampling magnetometry (VSM) measurements performed along the in-plane orientation. While pure SNO is not magnetic at room temperature, the annealed SNON samples all show clear magnetic hysteresis (Fig. 4a). Although the magnetization is enabled by the introduction of N, it weakens as the N concentration further increases. As shown in Figure 4a and 4b, the measured saturation magnetization decreases by around 50% as the N concentration increases from $x = 0.3$ to $x = 0.6$. Meanwhile, the coercive field becomes larger at higher N concentrations.

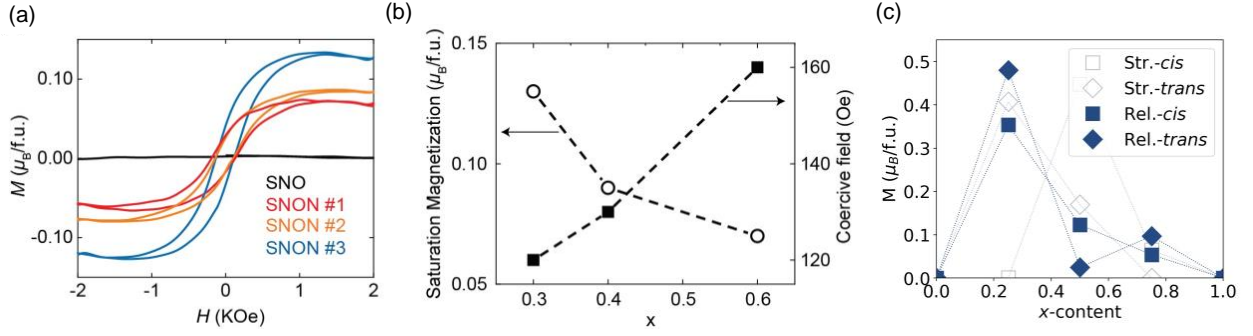

*Figure 4: Magnetic properties of SrNbO$_{3-x}$N$_x$ films. (a) The magnetic hysteresis curves experimentally measured on four films with different N contents. (b) In films annealed in NH$_3$, as the content of N increases, the saturation magnetization decreases and the coercive field increases. (c) Theoretically obtained magnetic moment per unit cell as a function of the nitrogen x-content for the cis- and trans-type configurations.*

To unveil the source of the magnetic behavior, we built a model from first-principles simulations, starting from the ground state crystal phase of SrNbO$_3$ ($x = 0$ in SrNbO$_{3-x}$N$_x$) that belongs to the orthorhombic *Pnma* symmetry space group[25] with an octahedral rotation pattern denoted by $a^-a^-c^+$ in Glazer notation[26]. When SrNbO$_3$ is exposed to in-plane strain (i.e., beyond -2.1%), our theoretical calculations indicate that its symmetry group and octahedral rotation goes to *I4/mcm* and $a^0a^0c^-$, respectively (See Fig, 2S in the Sup. Mat). We also find that in SrNbO$_2$N ($x = 1$ in SrNbO$_{3-x}$N$_x$), the incorporation of N suppresses the $a^-$ and $a^+$ rotations and results in a structure with the same *I4/mcm* symmetry group and $a^0a^0c^-$ rotation pattern in the absence of strain (see Figure 1a). This structural phase has been reported in the literature for temperatures $T > 300K$ for SrNbO$_2$N[27]. However, it is worth noticing that this space group is assigned considering indistinguishable N/O occupations in the corned octahedral sites. Given these results in our electronic structure computations, we took the SrNbO$_3$ structure in the *I4/mcm* symmetry group to be the parent structure for the N substitutions. From the parent material we replaced the corresponding number of oxygen atoms by nitrogen to realize the different N-concentrations, i.e. $x =$0, 0.25, 0.5, 0.75, and 1.0 for SrNbO$_{3-x}$N$_x$. Then for each of the structures we fully optimized the geometry to arrive at the ground state configurations for each N concentration.

To theoretically investigate the ground state magnetic properties for each value of *x,* we calculated the total energy of the A-, C- and G-type antiferromagnetic (AFM) ordering as well as the ferromagnetic state. When $x = 0$ (corresponding to the parent material SrNbO$_3$), the magnetic moment per Nb atom is about $m = 0.815\mu_B$, and the magnetic ground state is A-type antiferromagnetic. Experimentally, no saturation magnetization was measured in the SrNbO$_3$ thin films, which agrees with our theoretical results. As the N concentration increases in our model, the magnetic moment per formula unit, µ$_B$/f.u. (each formula unit hosts 5 atoms while our unit cell contains 20 atoms for the considered SrNbO$_{3-x}$N$_x$ structure.) first increases, reaching a maximum when the nitrogen concentration is between $x = 0.2$ and $x = 0.4$, as shown in Figure 4c. For higher concentrations, the magnetic moment per formula unit decreases and finally vanishes at $x = 1.0$. More careful analysis of our calculations reveals that the nonmonotonic behavior of the magnetic moment corresponds to a nontrivial magnetic phase transition as a function of the nitrogen concentration. As the nitrogen content increases, the computed magnetic ground state undergoes a transition from A-type AFM state at $x = 0$ to a ferromagnetic state at $x = 0.25$. The ferromagnetic state remains for $x = 0.5$ and 0.75, before arriving at a nonmagnetic state for $x = 1.0$, where the Nb ions have the valence state of +5 with $4d^0$ configuration. The effect of the strain imposed by the substrate was also considered in our analysis. In Figure 4c, we compare the results from calculations with and without strain for the lowest energy *cis*-type and *trans*-type configurations for each concentration. The epitaxial strain clearly shifts the nitrogen concentration corresponding to the maximum magnetic moment to greater value, but the overall behavior is not altered. Consequently, both models fairly reproduce and explain the experimentally found magnetic behavior condensed in Figure 4a and 4b.

As is widely known, ferromagnetism in complex oxides can arise from the double exchange interaction as a consequence of the incorporation of multiple oxidation states. Since each compound in the SrNbO$_{3-x}$N$_x$ series host a ratio of the Nb oxidation states of the two stochiometric endpoints, double exchange is a plausible mechanism for the emergent magnetism. However, the anisotropy and screening seen in the exchange parameters (See Figure 2S-6S) suggest an alternative or cooperative mechanism. Specifically, the exchange parameters are an order of magnitude larger along the *c*-axis than the parameters within the *a-b* plane for each structure. Likewise, the exchange parameters are sensitive to the local chemical environment as seen in the distinction between the interactions across the Nb-O-Nb and N-Nb-N bond chains, as well as in the reduced magnetization that accompanies the $a^+a^+c^0$ rotations in the $x = 0.75$ *cis*-type structure in contrast with the $a^0a^0c^-$ *trans*-type structure with the same stoichiometry (See Figure 5S-6S). Qualitatively, these factors suggest that the magnetism may be influenced by the orbital overlap in addition to the partial orbital filling.

Performing an analysis with the LOBSTER[28-30] code, we obtained some insights into the nature of the orbital overlap between Nb-O and Nb-N pairs (See Figure 7S**).** The Crystal Orbital Hamilton Population (COHP)[31] analysis allows the visualization of the accumulation (positive values) or depletion (negative values) of the charge density between the pairs of atoms compared with nonbonding atomic orbitals. The COHP reveals that the incorporation of N is accompanied by a significant reduction in the charge density between the Nb 4$d$ orbitals and the O 2$p$ orbitals in each of the magnetic cases. The depletion indicates that the Fermi level states are antibonding. Fermi level antibonding is an instability that is generally alleviated through structural distortions or though charge redistribution that results in an itinerant magnetic moment[32]. In each of the ferromagnetic $SrNbO_{3-x}N_x$ structures, the antibonding is reduced by an energy penalty placed on one of the spin channels, driving exchange splitting. Itinerant magnetism is further supported by correlating the size of the exchange splitting with the spontaneous magnetization. As the nitrogen content increases, the fermi level lowers in energy. Consequently, the antibonding at the fermi level reduces as the fermi level moves towards the tail of the Nb 4d energy levels. Since there is less antibonding intrinsically, the exchange splitting required reduces with increasing nitrogen content, tracing the evolution of the magnetization. Additionally, the $a^+a^+c^0$ rotations in the $x = 0.75$, *cis*-type structure appear to mitigate the antibonding in competition with the itinerant mechanism, reducing the magnetization compared with the *trans*-type structure with the same nitrogen content. Moving forward, X-ray magnetic dichroism can be employed in investigations aiming to explore the magnetic behavior of these oxynitrides in detail and to define the possible existence of a magnetic structure in $SrNbO_3$ thin films unambiguously.

*Ferroelectric properties:*

Pure SNO is centrosymmetric. Substituting oxygen with nitrogen breaks the local symmetry and can lead to the onset of spontaneous electric polarization. Figure 5a shows the piezo response force microscopy (PFM) images measured from the three SNON samples. Before imaging, a standard box-in-box biased contact scan procedure was carried out. Here, a 2 µm square is first scanned with a probe bias of 10 V. Then the smaller 1 µm square region at the center is scanned for a second time with a probe bias of -10 V. Persistent ferroelectric switching features are observed in SNON #3, where the regions scanned with opposite biases exhibit maximized PFM magnitudes and a relative phase shift of π, consistent with the formation of opposite residual polarizations. In comparison, such effects are much weaker in SNON #2 and completely vanish in SNON #1. Similar to the magnetic ordering, the ferroelectricity in these samples requires the N but is strongest when the N concentration is small.

According to our simulations, the polar behavior exists in both *cis*- and *trans*-type configurations. Moreover, it is worth noticing that it is advantageous to work with the $SrNbO_{3-x}N_x$ crystals compared to other oxynitrides, because the energy difference between the *cis*- and *trans*-type configurations is found to be approximately $\Delta E \approx 65 - 170$ meV/f.u., depending on the nitrogen content. This energy difference is remarkably smaller than the values found in other oxynitrides such as $BaTaO_2N$ ($\Delta E \approx 300$ meV/f.u.)[12,15] and $SrTaO_2N$ ($\Delta E \approx 200$ meV/f.u.)[33]. Hence, the smaller energy difference would allow the experimental realization of coexisting *cis* and *trans* domains $SrNbO_{3-x}N_x$ in thin films, where the strain from the substrate may help stabilize the energetically unfavored *trans*-type structures.

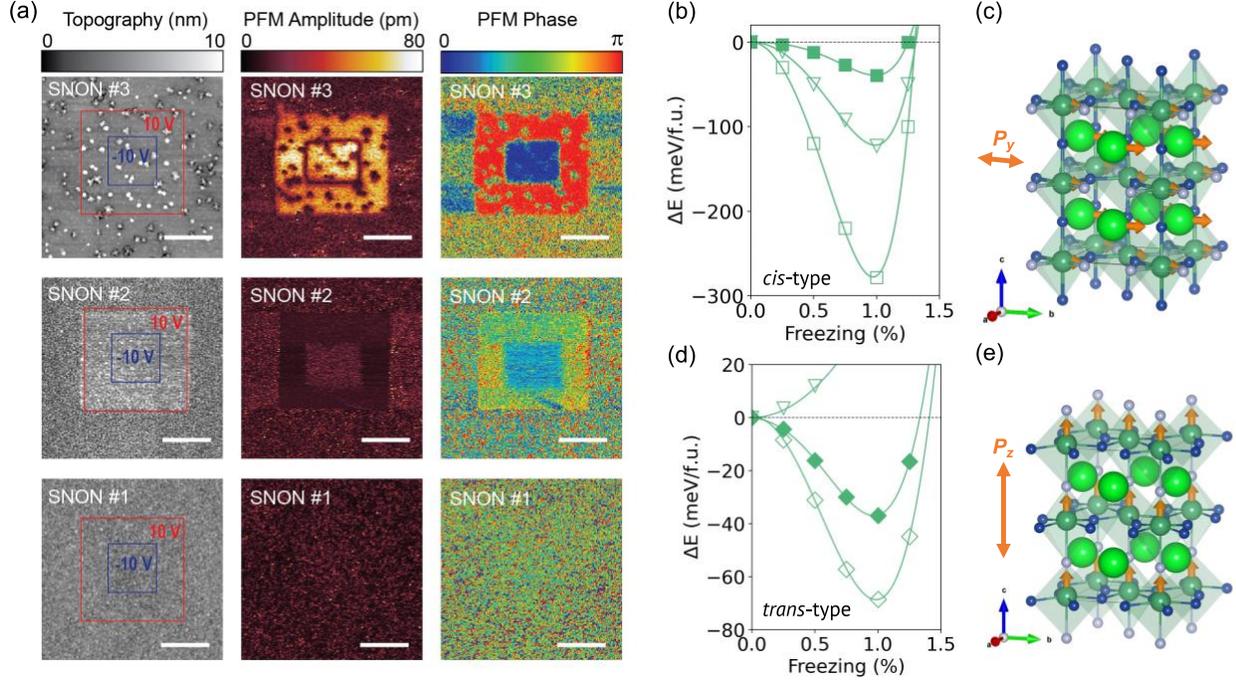

*Figure 5: Ferroelectric response in SrNbO$_{3-x}$N$_x$ (a) Piezo response force microscopy images measured on the three annealed samples. Before the PFM imaging, an identical box-in-box biased AFM writing sequence is applied to all three samples, which is shown in the topography images using red and blue lines. Persistent polarization switching effects are only observed in the sample with the lowest N concentration (SNON #3). Switchable polar distortion and the computed double-well energy curves obtained by freezing or condensing (i.e. introduction the atomic displacements associated with a particular phonon mode) the $\Gamma_3^-$ mode for the cis-type, at (b) and (c) and at the trans-type structures, at (d) and (e). For the cis-type structure, the transition goes from the Pmma to the Pmc2$_1$ polar space group. For the trans-type anionic ordering, the distortion leads the structure from the I4/mcm to the I4cm (No. 108). Here, the relaxed, out-of-plane, and in-plane strained data is denoted by filled squares, empty squares and triangles, respectively.*

Based on the analysis of the phonon-dispersion curves, the existence of the polar *Pmc2$_1$* symmetry group (No. 26) was identified in the *cis*-type anionic configuration[34]. In this phase, the octahedral rotations couple to the Sr displacements driving the ferroelectric distortion in this particular anionic ordering, see Fig. 5b and 5c. We find that strain applied both out-of-plane (i.e. the N-Nb-N zig-zag chains are aligned in the plane) as well as in-plane (i.e. the zig-zag chains are aligned out of plane) enhances the polar distortion considerably, denoted by the energy profiles shown in Figure 5b. In the out of plane case, the symmetry group remains *Pmc2₁* whereas for the in-plain strain case, the symmetry is lowered to the *Pm* (No. 6) group. When comparing the unstrained and strained SrNbO₂N, we observed a gain in energy of 238 meV/f.u and 83 meV/f.u for the out-of plane and in-plane strain, respectively, as observed from Fig 5b.

Regarding the *trans*-type configuration of SrNbO₂N, the vibrational landscape also reveals the mechanism for the polar ground state. Shown in Figure 5d, a $\Gamma_3^-$ phonon condenses into the *I4/mcm* space group (No. 140), leading the structure to the polar *I4cm* (No. 108) space group. This $\Gamma_3^-$ mode involves the off-centering of the Nb sites inside the NbO₄N₂ octahedra, where the octahedral rotations and the polarization coexist, as shown in Figure 5e. Interestingly, the energy-well curves computed in each case (see Figures 5d and 5e) demonstrate an energy increase of 37 and 68 meV/f.u as a result of the condensed mode for the unstrained and out-of-plane strained films, respectively. These energies indicate a tangible spontaneous polarization in the oxynitride compound, where the strain enhances the polar distortion. When the strain from STO is applied in-plane, the symmetry reduces from *I4cm* (No. 108) to *Fmmm* (No. 69) where the polar Nb distortions are not allowed as suggested by the single-energy well presented in Fig 5d.

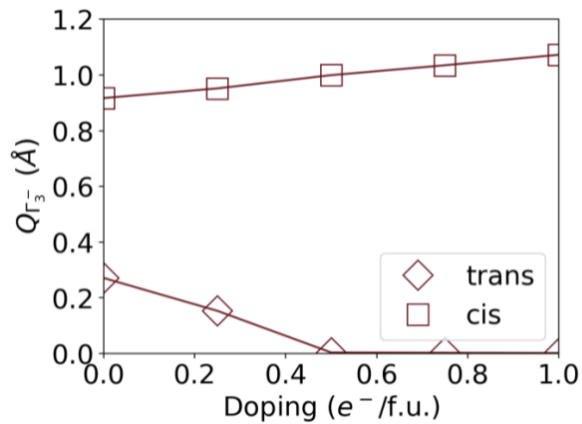

*Figure 6: Ferroelectricity and electron doping in SrNbO₃₋ₓNₓ. Polar mode contribution to the overall distortion as a function of the electronic doping with out of plane epitaxial strain is presented for both anionic orderings.*

Finally, we studied the effect of electron doping on the polar distortions of the *cis*- and *trans*-type anionic orders in the presence of out-of plane epitaxial strain in order to explore regions with equivalent low nitrogen concentrations. Since the O:N ratio changes the Nb oxidation state from +4 to +5, the electronic reconstruction results in partial charge occupation of the Nb-4$d$ level. As such, for nitrogen contents close to $x = 1$, the Nb exhibits a $4d^0$ configuration, whereas, for $x = 0$ a $4d^1$ is expected in agreement with our results. Thus, electron doping of 0.00, 0.25, 0.50, 0.75, and 1.00 e/f.u. corresponds to $x = 1.00, 0.75, 0.50, 0.25 \wedge 0.00$, respectively. This analysis helps us to understand and explain the existence of the polar distortion and magnetic response close to $x = 0.30$. In Figure 6, we show that the *cis*-type ordering is enhanced, despite expecting the polar distortion to vanish due to the electron doping in $SrNbO_{3-x}N_x$. In contrast, the polar distortion is destroyed in the *trans*-type ordering in agreement with the mechanism that drives the Nb ions off-center in this anionic configuration[2]. The survival of the polar distortion has also been observed in $BaTiO_3$[35] and $Bi_2WO_6$[36]. Moreover, the experimental feasibility of the ferroelectric switching has also been demonstrated for polar metals[37-39]. These results suggest that polar distortions and magnetism can coexist in this oxynitride. Therefore, based on our calculations and analysis, we attribute the measured PFM response to the coexistence of both, *cis*- and *trans*-type configurations with ferroelectricity.

**Conclusions**

In conclusion, our theoretical and experimental analyses show the multiferroic nature of the multi-anion $SrNbO_{3-x}N_x$ thin films. Since the energy difference between *cis*- and *trans*-type structures is small, both configurations can coexist in our samples. The measured ferromagnetism is ascribed to both the *cis*- and *trans*-type configurations, and the ferroelectric responses come from the displacement of the Nb cations from the centers of the octahedra in the *trans*-type and the coupling between the octahedral rotations and Sr-displacements in the *cis*-type configuration. These results highlight that through multi-anion configurations and epitaxial strain, a delicate balance, between the partial *transition-metal-d state* occupation, responsible for the magnetic response, and the cation site distortions, responsible for ferroelectric polarization, can be tuned to design multiferroism.

**Acknowledgements**

This work has been supported by the grants NSF SI2-SSE 1740112, DMREF-NSF 1434897, DOE DE-SC0016176 and DE-SC0019491. The theoretical results were obtained thanks to the XSEDE facilities which are supported by the National Science Foundation under grant number ACI-1053575. The authors also acknowledge the support from the Texas Advances Computer Center (with the Stampede2 and Bridges supercomputers), the OFFSPRING PRACE project (using the DECI resource BEM based in Poland at Wrocław) and on the CECI facilities funded by F.R.S-FNRS (Grant No. 2.5020.1) and Tier-1 supercomputer of the Fédération Wallonie-Bruxelles funded by the Walloon Region (Grant No. 1117545). A.C. Garcia-Castro acknowledge the grant No. 2489 entitled "Investigación y predicción de propiedades ferroeléctricas, magnéticas y magnetoeléctricas de oxinitruros tipo $Sr(Nb,Ta)O_{3-x}N_x$" supported by the VIE – UIS. Also, we thank the support from the GridUIS-2 experimental testbed, developed under the Universidad Industrial de Santander (SC3-UIS) High Performance and Scientific Computing Centre, with support from UIS Vicerrectoría de Investigación y Extensión (VIE-UIS) and several UIS research groups as well as other funding resources.

**Methods**

*Sample Preparation:*

Prior to the growth experiments, as-received $SrTiO_3$ (001) substrates were first exposed to UV ozone illumination for 20 minutes then cleaned with acetone and methanol and rinsed with DI water. After cleaning, the substrates were etched by buffered-HF (BHF) for 60 seconds to make the surface $TiO_2$-terminated. After chemical treatment, STO substrates were annealed in a tube furnace at 1050 °C for 5 hours with the oxygen flow rate of 500 sccm. The regular terraces with the width of ~ 200 nm (the corresponding miscut angle is about 0.11°) were found by atomic force microscopy after thermal annealing. The deposition of $SrNbO_3$ (SNO) thin films were carried out by PLD at 730 °C with the oxygen background pressure of 1E-5 Torr. The $SrNbO_{3-x}N_x$ structures were created in a dedicated $NH_3$ annealing station, where the annealing temperature, the $NH_3$ flow rate and time can all be programmed. In our work, we kept the annealing temperature and time at 950 °C and 3 hours, respectively. By varying the $NH_3$ flow rate from 29 sccm to 100 sccm, we tuned the nitrogen content in $SrNbO_{3-x}N_x$ from $x = 0.3$ to $x = 0.6$.

*Magnetism and piezoelectricity measurements:*

The ferromagnetic response was measured by Physical Properties Measurement System (PPMS) using the VSM insert. The piezo force microscopy was performed by Asylum MFP-3D system at ambient conditions.

*DFT calculations:*

We performed first-principles calculations within the density-functional theory, DFT[40,41], as implemented in the VASP code (version vasp5.3.3)[42,43]. We used the projected augmented wave (PAW) method[44] to represent the valence and core electrons. The electronic configurations considered as valence electrons were: Sr ($4s^24p^65s^2$, version 07Sep2000), Nb ($4p^64d^45s^1$, version 08Apr2002), N ($2s^22p^3$, version 08Apr2002), and O ($2s^22p^4$, version 08Apr2002). The exchange-correlation was represented within the generalized gradient approximation (GGA) with the PBEsol approximation[45]. The magnetic character was considered, and the *d*-electrons were corrected by means of the DFT+U within the Liechtenstein formalism[46] with a U value of 4.0 eV. The periodic solution of these crystalline structures was represented by using Bloch states with a Γ-centered *k*-point mesh of 8x8x6 and 600 eV of energy cut-off, which has been tested already to give forces convergence to less than 0.001 eV·Å$^{-1}$. With the aim to treat different O and N concentrations the site occupation disorder approach was employed with the SOD code implementation[47]. To do so, the SrNbO$_{3-x}$N$_x$ stoichiometric relationship was considered in which, x = 0, 0.25, 0.50, 0.75, and 1.00 substitutions were made. As a starting symmetry, we used the *I4mcm* structure demonstrated to be the ground state when *x* = 1, i.e. SrNbO$_2$N[48]. Although the SrNbO$_3$, i.e. *x* = 0, is reported to be *Pnma* ($a^-a^-c^+$) structure in its ground state, under strain an *I4/mcm* ($a^oa^oc^-$) is calculated. Finally, the phonon-dispersion curves were computed based on the DFPT approach[49] as implemented in the VASP code. The exchange parameters were calculated using the TB2J[50] python package, which calculates the exchange integrals directly from a projected tight binding Hamiltonian[51, 52]. The necessary tight binding projections were calculated using Wannier90[53] by projecting over the Nb 4*d* orbitals as well as the O and N 2*p* orbitals. The spread functional was considered to be converged when the difference between iterations was less than 10$^{-10}$. Finally, the electron doped calculations were performed at fixed cell parameter of the undoped case[54]